\documentclass[11pt,a4paper,english,showpacs,superscriptaddress,prd,aps,preprint]{revtex4-1}
\usepackage{amssymb}
\usepackage{amsmath}
\usepackage{graphicx}
\usepackage{graphics}
\usepackage{amsmath}
\usepackage{dcolumn}
\usepackage{amssymb}
\usepackage{bm}
\usepackage[latin1]{inputenc}
\usepackage{hyperref}
\usepackage{babel}
\newcommand{\bea}{\begin{eqnarray}}
\newcommand{\eea}{\end{eqnarray}}

\newcommand{\pa}{\partial}

\newcommand{\be}{\begin{equation}}
\newcommand{\ee}{\end{equation}}

\begin{document}

\title{K\"ahlerian Effective Potentials for Chern-Simons-Matter Theories }

\author{J. M. Queiruga}
\email{queiruga@if.usp.br}
\affiliation{Instituto de F\'\i sica, Universidade de S\~ao Paulo\\
Caixa Postal 66318, 05315-970, S\~ao Paulo, S\~ao Paulo, Brazil}

\author{A. C. Lehum}
\email{andrelehum@ect.ufrn.br}
\affiliation{Escola de Ci\^encias e Tecnologia, Universidade Federal do Rio Grande do Norte\\
Caixa Postal 1524, 59072-970, Natal, Rio Grande do Norte, Brazil}

\author{Mir Faizal}
\email{lf2mir@uwaterloo.ca}
\affiliation{Department of Physics and Astronomy,\\
University of Waterloo, Waterloo,\\
Ontario N2L 3G1, Canada}
\affiliation{Department of Physics and Astronomy,\\
University of Lethbridge,\\
Lethbridge, Alberta, T1K 3M4, Canada}

\begin{abstract}
In this paper, we will calculate the effective potential for a theory of multiple M2-branes.
As the theory of multiple M2-branes can be described by a Chern-Simons-matter theory, this will be done by 
calculating the K\"ahlerian effective potential for a Chern-Simons-matter theory. This calculation will 
be performed in $\mathcal{N} = 1$ superspace formalism. We will initially study an Abelian Chern-Simons-matter 
theory, and then generalize those results to the full non-Abelian Chern-Simons-matter theory. We will obtain explicit expressions 
for the  superpropagators for this theory.  These superpropagators will be used to calculate the one-loop effective potential. 
\end{abstract}

\maketitle

\section{Introduction}

It is known that the action for multiple M2-branes should have $\mathcal{N} = 8$ supersymmetry. 
This is because the superconformal field theory describing multiple M2-branes is dual to the eleven dimensional supergravity 
on $AdS_4 \times S^7$. Furthermore, we have $AdS_4 \times S^7 \sim [SO(2,3)/ SO (1, 3)]\times  [SO(8)/ SO(7)] \subset OSp(8|4)/[SO(1,3)
\times SO(7)]$. Thus, the  supergroup $OSp(8|4)$ can get  realized as $\mathcal{N} = 8$ supersymmetry of the   field theory dual to the eleven 
dimensional supergravity  on $AdS_4 \times S^7$.  Furthermore, the  on-shell degrees of this theory are 
exhausted by bosons and physical fermions. So, the gauge sector of the  theory for multiple M2-branes cannot have any on-shell degrees of freedom. 
These requirements are met by a theory called  the   BLG theory \cite{1a}-\cite{5a}. This theory is a Chern-Simons-matter theory.  However,
the gauge fields in this theory are valued in a Lie $3$-algebra rather than a conventional Lie algebra. This theory describes two M2-branes, 
and it  is not possible to use the BLG theory to describe more than two M2-branes. This is because there is only one known example of finite dimensional 
Lie $3$-algebra, and this example describes two M2-branes. However, by complexifying the matter fields, the gauge sector of the BLG theory can be written as 
sum of two Chern-Simons theories, with levels $k$ and $-k$. The gauge fields of these Chern-Simons field theories are valued in regular Lie algebra, and the 
matter fields transform in the bifundamental representation. 

It is possible to relax the requirement of manifest $\mathcal{N} =8$ supersymmetry and generalize this approach to a Chern-Simons-matter theory with $\mathcal{N} =6$ 
supersymmetry. This theory is called ABJM theory  and it coincides with the BLG theory for the only known example of the Lie $3$-algebra \cite{apjm}-\cite{abjm1}. 
The gauge symmetry of this theory is represented by Chern-Simons theories with the  gauge group  $U_k (N) \times U_{-k}(N)$, where $k$ and $-k$ are Chern-Simons levels. 
It may be noted that even though this theory only has manifest $\mathcal{N} = 6$ supersymmetry, its supersymmetry is expected to be enhanced to the full 
$\mathcal{N} = 8$ supersymmetry for $k=1$ or $k=2$ \cite{mp}. This enhanced occurs due to the effects generated from monopole operators. As the ABJM theory 
has a gauge symmetry, we will need to fix a gauge for calculate the K\"ahlerian effective potential. This can be done by adding the gauge fixing and ghosts
terms to the original ABJM action. The gauge fixing of the ABJM theory and the BRST symmetry for this theory have been throughly studied \cite{br}-\cite{rb}.

In this paper, we will analyze the Chern-Simons-matter theories in    $\mathcal{N} = 1$ superspace. It may be noted that even though   this will only have 
manifest 
$\mathcal{N} = 1$ supersymmetry, the actual theory will have higher amount of supersymmetry. We will use $\mathcal{N} = 1$ superspace formalism since the K\"ahlerian 
effective potentials is well understood in $\mathcal{N} = 1$ superspace formalism \cite{1,Ferrari:2010ex,Lehum:2010tt,Jackiw,a,b,2}. 
It may be noted that the ordinary Chern-Simons theory does not get renormalized,  except for a  finite one-loop
shift \cite{s1}-\cite{s2}.  The renormalization of Chern-Simons theory coupled to matter fields has also been studied 
\cite{s3}. The  renormalization of supersymmetric Chern-Simons-matter theories has also been studied 
\cite{s4}-\cite{s6}.   The matter fields exist in the fundamental representation of the gauge group. 
However, in a theory of multiple M2-branes, the matter fields exist in the bifundamental representation of the gauge group. 
It is possible to express the action of two M2-branes as    matter-Chen-Simons theory where the gauge fields are valued in a Lie $3$-algebra, and the  
  one-loop renormalization such a theory has also been analyzed  
\cite{s7}-\cite{s8}. The  scattering amplitudes in the ABJM theory have also been studied \cite{s9}-\cite{s10}. 
However, it is important to study the correction to the K\"ahlerian effective potential. This is because we expect 
to understand the dynamics of M5-branes by studding the  M2-branes ending on M5-branes \cite{m9m2}-\cite{2f}. This analysis is done 
using the superspace formalism. So, we will need to understand the one-loop corrections to the K\"ahlerian effective potential, 
to understand the quantum behavior of this theory, Furthermore, certain symmetries can be broken in the theory of multiple M2-branes. 
In fact, the inclusion of a mass term breaks the conformal invariance of the ABJM theory \cite{1s}-\cite{2s}. So, we will need to compute 
the effective K\"ahlerian effective potential for various deformations of the ABJM theory. Even though we only compute 
the K\"ahlerian effective potential for the ordinary ABJM theory, the method used in this paper can be easily generalized to study 
the K\"ahlerian effective potential for various deformations of the ABJM theory. Thus, apart from explicitly demonstrating the fact that 
the K\"ahlerian effective potential does not get corrected at one-loop, we also develop a formalism which can be used for deriving  
such results for various deformations of the ABJM theory. 

The Lagrangian for the ABJM  theory can be written as 
$
\mathcal{L}=\mathcal{L}_M+\mathcal{L}_{CS}+\tilde{\mathcal{L}}_{CS}, 
$
where $\mathcal{L}_{CS}$ and $\tilde{\mathcal{L}}_{CS}$ represent  the Chern-Simons terms with levels $k$ and $-k$. Here $\Gamma^\alpha$ and $\tilde{\Gamma}^\alpha$
represent  the Lie algebra valued $\mathcal{N}=1$ spinor superfields which can be used to construct the field strength, 
$
W_\alpha=\frac{1}{2}D^\beta  D_\alpha \Gamma_\beta -\frac{i}{2}\lbrack  \Gamma^\beta ,D_\beta  
\Gamma_\alpha \rbrack-\frac{1}{6}\lbrack \Gamma^\beta ,\lbrace \Gamma_\beta ,\Gamma_\alpha\rbrace\rbrack$
and 
$
\tilde{W}_\alpha=\frac{1}{2}D^\beta  D_\alpha \tilde{\Gamma}_\beta 
-\frac{i}{2}\lbrack  \tilde{\Gamma}^\beta ,D_\beta  \tilde{\Gamma}_\alpha \rbrack-\frac{1}{6}\lbrack 
\tilde{\Gamma}^\beta ,\lbrace \tilde{\Gamma}_\beta ,\tilde{\Gamma}_\alpha\rbrace\rbrack.
$
The matter content of the $U(N)_{-k}\times U(N)_{k}$ ABJM theory can be represented by the superfields  $(\Phi^I)^a_{\, \hat{a}}$,
such that they transform in the bifundamental representation. The Lagrangian for the matter fields is given by a sum of the kinetic term with the potential term. 
The potential term is represented by the superpotential $\mathcal{V}$. The Lagrangian for the kinetic part is constructed from covariant derivatives, 
$
\nabla _\alpha \Phi ^I= D_\alpha \Phi^I+i\Gamma_\alpha \Phi^I-i \Phi^I \tilde{\Gamma}_\alpha, $ and $
\nabla _\alpha \Phi ^{I\dagger}= D_\alpha \Phi^{I\dagger}+i\tilde{\Gamma}_\alpha \Phi^{I\dagger}-i \Phi^{I\dagger} \Gamma_\alpha.
$

\section{Abelian Chern-Simons-Matter Theory }

In this section we will calculate the  K\"ahlerian effective potential for the Abelian Chern-Simons-matter theory in $\mathcal{N} = 1$ superspace formalism \cite{abel}-\cite{abel1}. This will be used to motivate the study of the K\"ahlerian effective potential for the full Chern-Simons-matter theory 
in the next section. The superfield strength $W_\alpha$ can now be written as 
$
W_\alpha=\frac{1}{2}D^\beta  D_\alpha \Gamma_\beta,    
\tilde{W}_\alpha=\frac{1}{2}D^\beta  D_\alpha \tilde{\Gamma}_\beta. 
$
Now we will analyze the effective potential for the following simple Lagrangian  
\begin{eqnarray}
\mathcal{L}=\int d^2\theta\, tr \left\{  \frac{k}{4\pi}\left(\Gamma^\alpha D^\beta 
D_\alpha \Gamma_\beta  -\tilde{\Gamma}^\alpha D^\beta  D_\alpha \tilde{\Gamma}_\beta \right)+\frac{1}{4}\nabla ^\alpha \Phi ^{I\dagger}\nabla _\alpha \Phi ^{I}  \right\}.
\end{eqnarray}

This Lagrangian can be used to understand the behavior of the K\"ahlerian effective potential for theories where the matter fields transform in bifundamental representation. It may be noted that all matter fields are complex and the superpotential of the Abelian ABJM theory can vanish. 
To find the effective potential we shift the scalar superfields as \cite{1}-\cite{2},
\begin{eqnarray} 
\Phi^I&\rightarrow&\frac{1}{\sqrt{2}}\left(\Phi^I+\phi_c^I  \right)\\
\Phi^{I\dagger}&\rightarrow&\frac{1}{\sqrt{2}}\left(\Phi^{I\dagger}+\phi_c^I  \right),
\end{eqnarray} 
where $\Phi^I,\Phi^{I\dagger}$ in the right  and side  are quantum complex superfields and $\phi^I_c$ are the real constant background superfields. 

In order to determine the K\"ahlerian effective potential, we will assume that, $\phi^I_c$ are constant and $D^\alpha \phi^I_c=0$. After the shifting the Chern-Simons part or the Lagrangian remains invariant (the VEV of the spinorial superfield must be zero, otherwise, we break the Lorentz symmetry). 
The matter part can be written as follows,
\begin{eqnarray} 
\mathcal{L}_M&=&\frac{1}{8}\int d^2 \theta \left\{ D^\alpha \Phi^{I\dagger} D_\alpha \Phi^I +i \Lambda^\alpha 
[D_\alpha \Phi^{I\dagger}\Phi^I-\Phi^{I\dagger}D_\alpha\Phi^I] \right.\nonumber\\
&&\left. +\Lambda^\alpha\Lambda_{\alpha}\Phi^{I\dagger}\Phi^I
 +2\Lambda^\alpha \Lambda_{\alpha}\phi_c^I\Sigma^I+2\phi_c^I\Lambda^\alpha D_\alpha \Pi^I+\Lambda^\alpha \Lambda_\alpha (\phi_c^I)^2 \right\},
\end{eqnarray} 
where
\begin{eqnarray}
\Lambda^\alpha=\Gamma^\alpha-\tilde{\Gamma}^\alpha,\quad \Sigma^I=Re[\Phi^I],\quad \Pi^I=Im[\Phi^I].
\end{eqnarray}

The  term $2\phi_c^I\Lambda^\alpha D_\alpha \Pi^I$ appears after shifting the matter fields. This term  contributes to the propagator of the fields 
$\Pi^I$ and $\Lambda^\alpha$. We can eliminate it with an adequate choice of the gauge fixing term. We use the following $R_\xi$ gauge-fixing term,
\begin{eqnarray}
\mathcal{L}_{gf}&=& \int d^2\theta \frac{1}{8\xi}\left\{  \left(D^\alpha \Gamma_\alpha+\xi \phi_c^I\Pi^I  \right)^2
     -\left( D^\alpha \tilde{\Gamma}_\alpha+\xi \phi_c^I\Pi^I  \right)^2  \right\}.
\end{eqnarray}
The Faddeev-Popov  term corresponding to this gauge fixing term is given by 
\begin{eqnarray}
\mathcal{L}_{gh}&=& \int d^2\theta \left\{ -\frac{1}{2}c^\dagger D^2 c-\frac{\xi}{4}\sum_I(\phi_c^I)^2c^\dagger c 
-\frac{\xi}{4}\phi_c^I c^\dagger \Sigma^I c \right. \nonumber \\ && \left.
+\frac{1}{2}\tilde{c}^\dagger D^2\tilde{c}+\frac{\xi}{4}\sum_I(\phi_c^I)^2\tilde{c}^\dagger \tilde{c}
+\frac{\xi}{4}\phi_c^I \tilde{c}^\dagger \Sigma^I \tilde{c}   \right\}.
\end{eqnarray}

Now adding this gauge fixing term and the ghost term to the original Chern-Simons-matter Lagrangian, we obtain the following superpropagators (see Appendix A), 
\begin{eqnarray} 
\langle T \Pi^I (p,\theta) \Pi^J (-p,\theta')\rangle&=&i\delta_{IJ}\frac{D^2}{p^2}\delta(\theta-\theta'),\nonumber \\
\langle T \Sigma^I (p,\theta) \Sigma^J (-p,\theta')\rangle&=&i\delta_{IJ}\frac{D^2}{p^2}\delta(\theta-\theta'),\nonumber \\
\langle T \Gamma_\alpha (p,\theta) \Gamma_\beta  (-p,\theta')\rangle&=&-\frac{i}{2}\left \{ \left(\frac{\pi}{2k~p^2}
-\frac{\sum_I(\phi^I)^2\xi^2}{2(p^2)^2}D^2\right)D_\alpha D_\beta \right. \nonumber\\
&&\left.+\left( \frac{\sum_I(\phi^I)^2\pi^2}{8k^2(p^2)^2}D^2-\frac{\xi}{p^2} \right)D_\beta  D_\alpha \right\}\delta(\theta-\theta'),\nonumber \\
\langle T \tilde{\Gamma}_\alpha (p,\theta) \tilde{\Gamma}_\beta  (-p,\theta')\rangle&=&\langle T \Gamma_\alpha (p,\theta) 
\Gamma_\beta  (-p,\theta')\rangle\vert_{k\rightarrow-k,\xi\rightarrow -\xi},\nonumber\\
\langle T \Gamma_\alpha (p,\theta) \tilde{\Gamma}_\beta  (-p,\theta')\rangle&=&-\frac{i}{2}\left\{\frac{\pi^2\sum_I 
(\phi^I)^2}{8k^2(p^2)^2} D^2D_\beta  D_\alpha+\nonumber\right.\\
&&\left.-\frac{\sum_I(\phi^I)^2\xi^2}{4(p^2)^2}D^2 D_\alpha D_\beta \right\}\delta(\theta-\theta'),\nonumber\\
\langle T c^\dagger (p,\theta) c (-p,\theta')\rangle&=& i\frac{D^2-\frac{\xi\sum_k\phi_k^2}{2}}{p^2+
\left(\frac{\xi\sum_k\phi_k^2}{2}\right)^2}\delta(\theta-\theta'),\nonumber\\
\langle T \tilde{c}^\dagger (p,\theta) \tilde{c} (-p,\theta')\rangle&=&- i\frac{D^2-\frac{\xi\sum_k
\phi_k^2}{2}}{p^2+\left(\frac{\xi\sum_k\phi_k^2}{2}\right)^2}\delta(\theta-\theta').
\end{eqnarray} 

The one-loop K\"ahlerian effective potential can  be written as 
\begin{eqnarray}
\Gamma_{1\,loop}=\frac{i}{2}\sum_i tr \log \mathcal{O}_i,
\end{eqnarray} 
where $\mathcal{O}_i$ is the operators acting in the quadratic part of the action. We will use the tadpole method to determine the one loop contribution. So, after integration with respect to $\theta'$ only terms proportional to $D^2$ will survive (since $D^2\delta(0)=1$), and after integration with respect to the internal momentum $p$, the only non vanishing contribution in the dimensional regularization scheme can arise from non zero poles in the propagators (since $\int d^dp\frac{1}{p^2}=0 $ in the dimensional regularization).  Therefore the one loop contribution of the fields $\Pi^I$, $\Sigma^I$, $\Gamma^\alpha$ and $\tilde{\Gamma}^\alpha$ vanishes, while the ghost contribution is compensated exactly with the antighost. So finally we obtain

\be
\Gamma_{1\,loop}=0
\ee

It is interesting to note that for the theory $U_k(1)\times U_{-l}(1)$, for $k\neq l$ the gauge loop contributes to the effective action even at one loop level with 

\be
\Gamma_{1\,loop}^{U_k(1)\times U_{-l}(1)}\propto \sum_I (\phi_c^I)^2\left( \frac{1}{k^2}+\frac{1}{l^2} \right)sgn \left(k-l\right)
\ee
which gives a zero for the theory $U_k(1)\times U_{-k}(1)$, but contributes non trivially for $k\neq l$.

\section{Non-Abelian Chern-Simons-Matter Theory}

In the previous section, we analyzed a simple Abelian Chern-Simons-matter theory in bifundamental representation. In this section, we will analyze the non-Abelian Chern-Simons-matter theory with gauge group $U(N)_k \times U(N)_{-k}$. The Lagrangian for the   gauge sector of this theory can be written as $\mathcal{L}_{CS} + \tilde{\mathcal{L}}_{CS}$, where
\begin{eqnarray} 
\mathcal{L}_{CS}&=&\frac{k}{2\pi}\int d^2 \theta\, tr\left\{\Gamma^\alpha
W_\alpha+\frac{i}{6}\lbrace \Gamma^\alpha,\Gamma^\beta\rbrace D_\beta \Gamma_\alpha 
 +\frac{1}{12}\lbrace \Gamma^\alpha,
\Gamma^\beta\rbrace\lbrace \Gamma_\alpha,\Gamma_\beta \rbrace    \right\},\\
\tilde{\mathcal{L}}_{CS}&=&-\frac{k}{2\pi}\int d^2\theta\,tr \left\{\tilde{\Gamma}^\alpha 
\tilde{W}_\alpha+\frac{i}{6}\lbrace \tilde{\Gamma}^\alpha,\tilde{\Gamma}^\beta \rbrace D_\beta  \tilde{\Gamma}_\alpha 
+\frac{1}{12}\lbrace \tilde{\Gamma}^\alpha,\tilde{\Gamma}^\beta \rbrace\lbrace \tilde{\Gamma}_\alpha,\tilde{\Gamma}_\beta \rbrace    \right\}.
\end{eqnarray} 
 
The Lagrangian for the matter sector of this theory can be written as   
\begin{eqnarray}
\mathcal{L}_M=\frac{1}{4}\int d^2 \theta \, tr \left\{ \nabla ^\alpha \Phi ^{I\dagger}\nabla _\alpha \Phi ^{I} 
+\mathcal{V} \right\},
\end{eqnarray}
where the covariant superderivatives are defined by
\begin{eqnarray} 
\nabla _\alpha \Phi ^I&=& D_\alpha \Phi^I+i\Gamma_\alpha \Phi^I-i \Phi^I \tilde{\Gamma}_\alpha\\
\nabla _\alpha \Phi ^{I\dagger}&=& D_\alpha \Phi^{I\dagger}+i\tilde{\Gamma}_\alpha \Phi^{I\dagger}-i \Phi^{I\dagger} \Gamma_\alpha,
\end{eqnarray} 
and the superpotential term $\mathcal{V}$ is given by
\begin{eqnarray}
\mathcal{V}=\frac{16\pi}{k}\epsilon^{IJ}\epsilon_{KL}\, \left(\Phi_I \Phi^{K\dagger}\Phi_J \Phi^{L\dagger} \right).
\end{eqnarray}
The matter content of this theory consists of two $N\times N $ matrices of  $\mathcal{N} =1$ superfields $(\Phi^I)^a_{\, \hat{a}}$ and their adjoints \cite{mp}. 

Now we will shift the superfields as follows,
\begin{eqnarray} 
\Phi^I&\rightarrow&\Phi^I+\rm{diag}(\phi^I_1,\phi^I_2,...,\phi^I_N)\\
\Phi^{I\dagger}&\rightarrow&\Phi^{I\dagger}+\rm{diag}(\phi^I_1,\phi^I_2,...,\phi^I_N)\ .
\end{eqnarray} 
where the diagonal matrices $\rm{diag}(\phi^I_1,\phi^I_2,...,\phi^I_N)$ are real classical superfields. It may be noted that again these matrices conform the full moduli space of the theory since the superpotential is identically zero. The shifted superpotential can  be written as
\begin{eqnarray}
\mathcal{V}_s=\mathcal{V}+\mathcal{V}_3+\frac{16\pi}{k}\left( \phi_j^1\phi_l^2 \left( \Phi^1_{lj}\Phi^2_{jl}- \Phi^1_{jl}\Phi^2_{lj}+ \Phi^{1\dagger}_{jl}\Phi^{2\dagger}_{lj}- \Phi^{1\dagger}_{lj}\Phi^{2\dagger}_{jl}\right)   \right),
\end{eqnarray}
where $\mathcal{V}_3$ contains 3-vertex terms. Now we can write the following expression for the remaining part of the matter sector, 
\begin{eqnarray} 
&&\frac{1}{4}\int d^2\theta \left \{ D^\alpha \Phi^{I\dagger}_{ij}D_\alpha \Phi^I_{ji} 
+i\left(D^\alpha \Phi^{I\dagger}_{ij}\Gamma_{\alpha\, ji}\phi^I_i-h.c. \right)
-i\left(D^\alpha \Phi^I_{ij}\tilde{\Gamma}_{\alpha\, ji}\phi^I_i-h.c. \right)\right.\nonumber\\ 
&&\left.   +	\phi^I_i 	\phi^I_i   \Gamma^\alpha_{ij}\Gamma_{\alpha\, ji} 
+\phi^I_i 	\phi^I_i  \tilde{ \Gamma}^\alpha_{ij}\tilde{\Gamma}_{\alpha\, ji} 
-2 \phi^I_i\phi^I_j \Gamma^\alpha_{ij}\tilde{\Gamma}_{\alpha\, ji}   \right \}+(\rm{interactions}).
\end{eqnarray} 

Now we use a Lorentz gauge fixing for calculating the effective potential, 
\begin{eqnarray}
\mathcal{L}_{gf}=\frac{1}{4\xi}\int d^2 \theta\, tr\left(D^\alpha \Gamma_{\alpha}\right)^2-\frac{1}{4\xi}\int d^2 \theta\, tr\left(D^\alpha \tilde{\Gamma}_{\alpha}\right)^2 ,
\end{eqnarray}
The corresponding Faddeev-Popov term corresponding to this gauge fixing term can be written as 
\begin{eqnarray}
\mathcal{L}_{FP}=i\int d^2 \theta\, Tr c^\dagger D^\alpha \nabla_\alpha c-i\int d^2 \theta\, Tr \tilde{c}^\dagger D^\alpha \nabla_\alpha\tilde{c}.
\end{eqnarray}
After  shifting the superfields,  the quadratic part of the action can be written as follows
\begin{eqnarray} \left( \begin{array}{cccccc}
\Phi^1_{ij} &\Phi^2_{ij}&\Phi^{1\star}_{ij}&\Phi^{2\star}_{ij}& \Gamma^\alpha_{ij}&\tilde{\Gamma}^\alpha_{ij} \\
\end{array} \right)
\left( \begin{array}{cc}
A &B  \\
C&D\\

\end{array} \right)
\left( \begin{array}{c}
\Phi^{1\star}_{kl}  \\
\Phi^{2\star}_{kl}  \\
\Phi^1_{kl}  \\
\Phi^1_{kl} \\
 \Gamma^\beta_{kl} \\
\tilde{\Gamma}^\beta_{kl}  
\end{array} \right).
\end{eqnarray}
The matrix operators $A,B,C$ and $D$ are suitable defined. So, the matrix operator $A$ is defined by 
\begin{eqnarray}A=
\delta_{il}\delta_{jk}\left( \begin{array}{cccc} 
-D^2&&&m(i,j)  \\
&-D^2&m(i,j)&\\
&m(i,j)&-D^2&	\\
m(i,j)&&&-D^2\\
\end{array} \right),
\end{eqnarray}
where  $m(i,j)=\frac{4\pi}{k}\left( \phi^1_j \phi^2_i - \phi^2_j \phi^1_i   \right)$. The matrix operator $D$ is defined by 
\begin{eqnarray}D=
\delta_{il}\delta_{jk}\left( \begin{array}{cc} 
D_{1\alpha \beta}&
-\frac{1}{2}\phi^I_j\phi^I_i  \\
-\frac{1}{2}\phi^I_j\phi^I_i &D_{2\alpha \beta  } \\
\end{array} \right),
\end{eqnarray}
where 
\begin{eqnarray}
 D_{1\alpha \beta} = \frac{k}{2\pi}D_\beta  D_\alpha+\frac{1}{2\xi}D_\alpha D_\beta +\frac{1}{2}C_{\beta \alpha}(\phi^I_j)^2, 
 \nonumber \\
 D_{2\alpha \beta  } = -\frac{k}{2\pi}D_\beta  D_\alpha-\frac{1}{2\xi}D_\alpha D_\beta 
+\frac{1}{2}C_{\beta \alpha}(\phi^I_j)^2.
\end{eqnarray}
The matrix operator $B$ is defined by 
\begin{eqnarray}B=
\delta_{il}\delta_{jk}\left( \begin{array}{cc} 
-i D_\beta  \phi^1_i&i D_\beta  \phi^1_j  \\
-i D_\beta  \phi^2_i&i D_\beta  \phi^2_j\\
i D_\beta  \phi^1_j&-i D_\beta  \phi^1_i\\
i D_\beta  \phi^2_j&-i D_\beta  \phi^2_i\\
\end{array} \right).
\end{eqnarray} 
Finally, the matrix operator $C$ is defined by 
\begin{eqnarray}
 C=B^T\vert_{\beta \rightarrow\alpha}. 
\end{eqnarray}

Now it is trivial to see that if we take the shift $\phi^I_i=\phi^I_j,\,\forall i,j$ then $m(i,j)=0$. It is possible to invert these matrices but the propagators have a complicated  form in this general case. In order to simplify them we shift the superfields as follows
\begin{eqnarray} 
\Phi^I&\rightarrow& \Phi^I+\phi^I_c\mathcal{I}_{N\times N}\\
\Phi^{I\dagger}&\rightarrow& \Phi^{I\dagger}+\phi^I_c\mathcal{I}_{N\times N}.
\end{eqnarray} 
In the non-Abelian case this shift represents a non-trivial restriction of the full moduli space of the theory. The scalar superpropagator has the following form
\begin{eqnarray}
\langle T \Phi^I_{ij} (p,\theta) \Phi^{I\dagger}_{\bar{i}\bar{j}} (-p,\theta')\rangle
&=& i\frac{\frac{\pi}{2k}\phi^I_c\phi^I_c p^2+(p^2
+m(m-\frac{\pi}{2k}\phi^I_c\phi^I_c))D^2}{p^2(p^2+m^2)}\nonumber \\ && \times \delta_{i\bar{j}}\delta_{j\bar{i}} \delta(\theta-\theta'),
\label{p1}
\end{eqnarray}
where $m=\frac{\pi}{k}\phi^I_c \phi^I_c$. Now  for $I\neq J$, we have 
\begin{eqnarray}
\langle T \Phi^I_{ij} (p,\theta) \Phi^{J\dagger}_{\bar{i}\bar{j}} (-p,\theta')\rangle
=-i\delta_{i\bar{j}}\delta_{j\bar{i}}\frac{\pi}{2k}\frac{\phi^I_c\phi^J_c
\left(p^2-mD^2\right)}{p^2(p^2+m^2)}\delta(\theta-\theta'). \label{p2}
\end{eqnarray}
Finally, we can write 
\begin{eqnarray} 
\langle T \Phi^I_{ij} (p,\theta) \Phi^I_{\bar{i}\bar{j}} (-p,\theta')\rangle&=&i\delta_{i\bar{j}}\delta_{j\bar{i}}\frac{\pi}{2k}\frac{\phi^I_c\phi^J_c\left(p^2-mD^2\right)}{p^2(p^2+m^2)}\delta(\theta-\theta') \\
\langle T \Phi^{I\dagger}_{ij} (p,\theta) \Phi^{I\dagger}_{\bar{i}\bar{j}} (-p,\theta')\rangle&=&-i\delta_{i\bar{j}}\delta_{j\bar{i}}\frac{\pi}{2k}\frac{\phi^I_c\phi^J_c\left(p^2-mD^2\right)}{p^2(p^2+m^2)}\delta(\theta-\theta').\label{p3}
\end{eqnarray} 

We use the following identity to invert the matrices:
\begin{eqnarray} M^{-1}= \left( \begin{array}{ccc}
N_{11} & N_{12} \\ N_{21} 
 & N_{22} \end{array}  \right)\end{eqnarray} 
where 
\begin{eqnarray} M= \left( \begin{array}{ccc}
A &B \\
C &D\end{array}  \right),\end{eqnarray}
and 
\begin{eqnarray}
N_{11} &=& (A-BD^{-1}C)^{-1}, 
\nonumber \\ 
N_{12} &=& -(A-BD^{-1}C)^{-1} BD^{-1},  
 \nonumber \\
N_{21} &=& -D^{-1}C (A-BD^{-1}C)^{-1}, 
 \nonumber \\
 N_{22} &=& D^{-1}+ D^{-1} C(A-BD^{-1}C)^{-1}BD^{-1}. 
\end{eqnarray}

In this gauge the ghosts do not contribute and the spinor superpropagators have the following form, 
\begin{eqnarray} 
\langle T \Gamma_{\alpha ij} (p,\theta) \Gamma_{\beta  \bar{i}\bar{j}} (-p,\theta')\rangle
&=&-\frac{i}{2}\frac{\delta_{i\bar{j}}\delta_{j\bar{i}}}{p^2}\left \{\left(\frac{\pi}{k}
 +\frac{\pi^2 \phi^I_c\phi^I_c}{2k^2~p^2}D^2\right)D_\alpha D_\beta  \right.\nonumber \\
&&+\left.\left( \frac{\pi m^2}{2k (p^2+m^2)}
-\frac{\pi m~D^2}{2k (p^2+m^2)}  \right)
D_\alpha D_\beta  \right \}\delta(\theta-\theta'),\label{sp1}\nonumber \\ 
\langle T \tilde{\Gamma}_\alpha (p,\theta) \tilde{\Gamma}_\beta  (-p,\theta')\rangle&=&\langle 
T \Gamma_\alpha (p,\theta) \Gamma_\beta  (-p,\theta')\rangle\vert_{k\rightarrow-k},\label{sp2}\nonumber \\
\langle T \Gamma_{\alpha ij} (p,\theta) \tilde{\Gamma}_{\beta  \bar{i}\bar{j}} (-p,\theta')\rangle&=&
-\frac{i}{2}\frac{\delta_{i\bar{j}}\delta_{j\bar{i}}}{p^2}\left \{ \left(-\frac{\pi^2 \phi^I_c\phi^I_c}{2k^2~p^2}D^2D_\alpha D_\beta
 -\frac{\pi}{k}C_{\alpha\beta }D^2\right) \right.\nonumber\\
&&\left. - \left( \frac{\pi m^2 }{2k~(p^2+m^2)}-\frac{\pi m~D^2}{2k~(p^2+m^2)} 
\right)D_\alpha D_\beta  \right \}\delta(\theta-\theta'),\label{sp3}  
\end{eqnarray} 

Now because of the mixing between 
 $\Phi^I$ and $\Gamma^\alpha, \tilde{\Gamma}^\alpha$ in the quadratic action there is a non zero pole in the propagators. We will use the tadpole method again. We can write the explicit integral over the propagator matrix after differentiating with respect to some parameter (let us say $\omega$) of the theory,
 
 \be
 \pa_\omega \Gamma_{\rm{1\,loop}}=\frac{i}{2} Tr M^{-1}\pa_\omega M \label{1loopa}
 \ee
 The one loop effective action will be obtained after performing the operations in the right hand side of (\ref{1loopa}), and integrate with respect to $\omega$. After the integration with respect to $\theta'$ only terms proportional to $D^2$ survive. The integration over internal momenta leads us to the following expression:
 
 \be
 Tr M^{-1}=-\frac{\vert m\vert}{4\pi}+\frac{\vert m\vert}{4\pi}=0
\ee
 where the first term is the contribution of the scalar superfields $\Phi^I,\, \Phi^{I\dagger}$, which is exactly compensated 
 with the contribution of the spinorial superfields $\Gamma_\alpha,\, \tilde{\Gamma}_\alpha$.   So, the one-loop effective potential still vanishes for
 non-Abelian Chern-Simons-matters theories. 
 \begin{eqnarray}
\Gamma_{\rm{1\,loop}}=\frac{i}{2} Tr \log M=0.
\end{eqnarray}
 It may be noted that it was expected that the K\"ahlerian effective potential will not get corrected 
 at one-loop, based on the fact that Chern-Simons-matter theories in general do not get   renormalized,  except for a  finite one-loop shift \cite{s1}-\cite{s6}. 
 However, it was important to show this explicitly. In fact,  it is possible to deform the ABJM theory, 
 and such superspace calculations can be used for analysing the deformed ABJM theory. We will like to point out that even 
   though we have not calculated the effects of such deformations, the methods used in this paper 
 can also be used for analysing the one-loop  K\"ahlerian effective potential for the deformed ABJM theory.
 Another advantage of using the superspace formalism is that we can now understand the behavior of the ABJM theory under a general shifting 
 of fields. So, now we can infer that under a more general shifting would be given by 
\begin{eqnarray} 
\Phi^I&\rightarrow&\Phi^I+\rm{diag}(\phi^I_1,\phi^I_2,...,\phi^I_N)\\
\Phi^{I\dagger}&\rightarrow&\Phi^{I\dagger}+\rm{diag}(\phi^I_1,\phi^I_2,...,\phi^I_N),
\end{eqnarray} 
we obtain exactly the same result. It may be noted that at least from the superspace perspective, it might be possible to obtain a 
non-trivial contribution beyond one-loop. This is because the shifting generates new 3-vertices in the superpotential term ($\mathcal{V}_s$), 
$
\mathcal{V}_s = \mathcal{V}+\frac{32\pi}{k}\left \{ \phi^1_c 
\Phi_2\left( [\Phi^{2\dagger},\Phi^{1\dagger}]+[\Phi^{2\dagger},\Phi_1]  \right)
  +\phi^2_c \Phi_1\left( [\Phi^{1\dagger},\Phi^{2\dagger}]+[\Phi^{1\dagger},\Phi_2]  \right)\right \},\label{potchern}
$
\\where $
  \mathcal{V}=\frac{16\pi}{k}\epsilon^{IJ}\epsilon_{KL}\, \left(\Phi_I \Phi^{K\dagger}\Phi_J \Phi^{L\dagger} \right). $
 It would be interesting to analyze such corrections, and study the implications of such amplitudes for the ABJM theory.

\section{Conclusion}
In this paper, we have  analyzed a Chern-Simons-matter theory in $\mathcal{N} = 1$ superspace formalism. This was done for studding
the one-loop effective potential for a theory describing multiple M2-branes. We initially studied an Abelian Chern-Simons-matter theory, and then generalized those results to the full non-Abelian Chern-Simons matter theory. Thus, we first  fixed a gauge by adding a gauge fixing term and a ghost term, and then we shifted the superfields. It was possible to calculate the expression for one-loop effective potential in this Chern-Simons-matter theory. Thus, we were able to calculate expressions for superpropagators of this theory. Finally, we used these superpropagators to calculate the one-loop effective potential. The vanishing value of such a one loop correction is in compete agreement with the results for $N=6$ ABJM theory. The superpropagator structure described here provides a nice starting point to discuss, for example non perturbative effects, with the advantage that bot fermionic and bosonic degrees of freedom are taken into account in one single supergraph.

It  is possible for  strings to end on D-branes in string theory. Similarly, it is possible M2-branes to end on  other objects in M-theory.
These other objects can be M5-brane, M9-branes, and gravitational waves  \cite{m9m2}. It may be noted that a system of multiple M2-branes ending on 
two M9-brane is expected to generate $E_8 \times E_8$ symmetry. This occurs due to the existence of the gravitational anomaly, and this is similar 
to the Horava-Witten formalism  \cite{hwsetup}-\cite{hwsetup1}. It may be noted that it is possible to understand the physics of M5-branes by analyzing 
a system of M2-branes ending on M5-branes. This makes a system of open M2-branes ending on a M5-brane very interesting. In fact, 
the BLG model has been  used to motivate a novel quantum geometry on the M5-brane. This has been done by studding a system of 
M2-branes ending on M5-branes in presence of  a constant $C$-field 
\cite{d12}. In fact, the BLG action with Nambu-Poisson $3$-bracket has been identified with  the action of  M5-brane, in presence of a 
large world-volume $C$-field \cite{M5BLG}. It may be noted that by analyzing a single M2-brane ending on a M5-brane, it was possible to 
study  non-commutative string theory on the M5-brane world-volume \cite{NCS1}-\cite{NCS3}. Thus, it is important to understand the BLG 
theory and the ABJM theory, in presence of a boundary. It may be noted that the BLG theory has been studied in presence of a boundary 
\cite{f1}-\cite{1f}. In fact, even boundary effects for the  ABJM theory have been studied \cite{f2}-\cite{2f}. It was observed that the boundary
breaks half the supersymmetry of the original theory. Furthermore, 
the gauge invariance of the theory could only be preserved by introducing new degrees of freedom on the boundary. It will be interesting to
analyze the one-loop effective potential for the ABJM theory in presence of a boundary. The issue of higher loop corrections to ABJM theory 
in $N=1$ superspace is under current investigation.

\appendix
\section{ Superpropagators in the K\"ahlerian approximation}

All operators in the bosonic sector can be written in terms of six projectors:

\be
P_0=1,\quad P_1=D^2,\quad P_2=\theta^2,\quad P_3=\theta^\alpha D_\alpha,\quad P_4=\theta^2 D^2,\quad P_5=i\pa_{\alpha\beta}\theta^\alpha D^\beta
\ee

Therefore, the K\"ahlerian approximation corresponds to take the projectors not involving explicitly Grassmann variables. The table of composition for these operators is the following

\begin{center}
  \begin{tabular}{  l | c | r }

    $\circ$ & $P_0$ & $P_1$ \\ \hline
    $P_0$ & $P_0$ & $P_1$ \\ \hline 
    $P_1$ & $P_1$ & $\square P_0$ \\

  \end{tabular}
\end{center}

Now if the quadratic part of the action can be written as $\Phi \mathcal{O} \Phi^\dagger$,  the propagator for the bosonic fields can be calculated by imposing the condition

\be
\mathcal{O}\sum_i p_i P_i=1
\ee
where the functions  $p_i$ will depend in general on the parameters of the theory and momenta. In the spinorial sector the strategy is the same. In this case we need to introduce the bi-spinorial projectors. The full basis has $14$ elements (see for example \cite{gallegos}), but in the  K\"ahlerian approximation we need only four

\be
R_{i,\alpha\beta}=i\pa_{\alpha\beta}P_i,\quad S_{i,\alpha\beta}=C_{\alpha\beta}P_i
\ee
where the composition can be read from the table above. In this sector, if $\mathcal{O}_{\alpha\beta}$ is the operator acting in the quadratic part of the action, i.e. $\Gamma^\alpha \mathcal{O}_{\alpha\beta} \Gamma^\beta $, the propagator of the spinorial superfields can be calculated by imposing the condition

\be
\mathcal{O}_{\alpha\beta}\left( \sum_i r_i R_i^{\beta\gamma}+ \sum_i s_i S_i^{\beta\gamma}  \right)=\delta_{\alpha}^{\,\,\gamma}
\ee

\section*{Acknowledgments}
This work was partially supported by Funda\c{c}\~ao de Amparo \`a Pesquisa do Estado de S\~ao Paulo (FAPESP), Conselho Nacional de Desenvolvimento Cient\'{\i}fico e Tecnol\'{o}gico (CNPq) and Funda\c{c}\~{a}o de Apoio \`{a} Pesquisa do Rio Grande do Norte (FAPERN).

\end{document}